\documentclass{amia}
\usepackage{graphicx}
\usepackage[labelfont=bf]{caption}
\usepackage[superscript,nomove]{cite}
\usepackage{color}

\usepackage{calrsfs}
\DeclareMathAlphabet{\pazocal}{OMS}{zplm}{m}{n}

\newcommand{\Lb}{\pazocal{L}}

\begin{document}

\title{Deep CHORES: Estimating Hallmark Measures of Physical Activity Using Deep Learning}

\author{Mamoun T. Mardini, PhD$^{1}$, Subhash Nerella, MS$^{1}$, Amal A. Wanigatunga, PhD, MPH$^{2}$, Santiago Saldana, MS$^{3}$, Ramon Casanova, PhD$^{3}$, Todd M. Manini, PhD$^{1}$}

\institutes{
    $^1$University of Florida, Gainesville, Florida, USA; $^2$Johns Hopkins University, Baltimore, Maryland, USA; $^3$Wake Forest University, Winston-Salem, North Carolina, USA\\
}

\maketitle

\noindent{\bf Abstract}
\textit{Wrist accelerometers for assessing hallmark measures of physical activity (PA) are rapidly growing with the advent of smartwatch technology. Given the growing popularity of wrist-worn accelerometers, there needs to be a rigorous evaluation for recognizing (PA) type and estimating energy expenditure (EE) across the lifespan. Participants (66\% women, aged 20-89 yrs) performed a battery of 33 daily activities in a standardized laboratory setting while a tri-axial accelerometer collected data from the right wrist. A portable metabolic unit was worn to measure metabolic intensity. We built deep learning networks to extract spatial and temporal representations from the time-series data, and used them to recognize PA type and estimate EE. The deep learning models resulted in high performance; the F1 score was: 0.82, 0.81, and 95 for recognizing sedentary, locomotor, and lifestyle activities, respectively. The root mean square error was 1.1 (+/-0.13) for the estimation of EE}

\section*{Introduction}
It is well known that regular and sufficient amounts of physical activity (PA) have tremendous benefits on reducing the risk of common chronic diseases and enhancing mental health, wellbeing and quality of life. Globally, one out of four adults (almost 1.4 billion) do not meet the World Health Organization (WHO) PA recommendations \cite{1}. Recently, WHO has published an action plan to enhance PAs with a target of 15\% reduction in physical inactivity by year 2030 \cite{2}. Although the effect of fitness trackers on increasing PA has yet rarely explored, a few recent studies have shown that these trackers can potentially change individuals behavior and increase PA \cite{chia2019behavior,brickwood2019consumer}. Therefore, there is a need to build models that can accurately recognize PA type and intensity.

Historically, the adopted approach used to recognize PA type and to estimate energy expenditure (EE) relied on data collected from the hip position in standardized laboratory settings. The advantage of the hip position over other positions is the closeness to the body’s center of the mass. Therefore, it offers a convenient and accurate approach for capturing ambulatory activity \cite{attal2015physical}. Recently, however, the wrist position has become popular for collecting accelerometer data due to a rise in smartwatches, convenience, ability to capture sleep quality and enhanced compliance in research studies \cite{full2018validation,idc,kerr2017comparison,migueles2017accelerometer}. Despite the popularity of wrist-worn accelerometers, there is a paucity of models that are deemed viable for accurately assessing wrist PA. The use of data from the wrist position to recognize PA type and estimate EE is challenging due to its limitation in quantifying and capturing large lower limb movements and other lifestyle activities.

In this paper, we targeted two pervasive issues for using accelerometers: i) recognizing PA type, which is a classification problem and ii) estimating energy expenditure, which is a regression problem. To address these issues, previous research has used several machine learning algorithms including decision tree \cite{staudenmayer2015methods}, random forests \cite{staudenmayer2015methods, ellis2014identifying}, and bag-of- words \cite{kheirkhahan2017bag}. This approach is generally better than traditional statistical regression-based approaches, due to their ability to handle a high resolution data across three accelerometer axes and are able to extract and utilize non-linear relationships more efficiently than traditional statistical-based approaches. The current machine learning approaches follow standard time series analysis, where relevant features are aggregated (feature engineering) on sliding windows (bouts) of the raw data followed by classification or regression. While this approach has successfully discovered nonlinear relationships, it stops short in identifying hidden features because it flattens the temporality in the data by summarizing it to time- and frequency-domain features. This will limit the ability of the machine learning model to extract the temporal features from the time series data, which are important for capturing the transitions between activity types. Lastly, the current machine learning approaches rely on the selection of relevant features, which requires domain expertise and varies significantly among researchers.

For these reasons, we embrace the power of deep learning in this paper, because it learns embeddings from the raw accelerometry data and identifies a comprehensive feature representation without user input. As recent examples, convolutional neural networks (CNN) and recurrent neural networks (RNN) demonstrated important breakthroughs in image \cite{17} and speech recognition \cite{18} and word embeddings \cite{19}. Recognizing activity type is an ideal fit for CNN and RNN to appropriately utilize the spatial and temporal representations of the time series signals \cite{20,21}. Deep learning is able to process the data through an organized hierarchy of neural networks that can extract progressively non-linear and abstract features that can be used for classification. Data are processed throughout the layers, where the input of each layer is the processed output of the previous layer. This architecture of deep learning allows the network to generalize and become domain-invariant, which means that after learning a specific pattern from data, the deep network will be able to recognize it on different data sources.

\begin{figure}[h]
	\centering
	\includegraphics[scale=0.75]{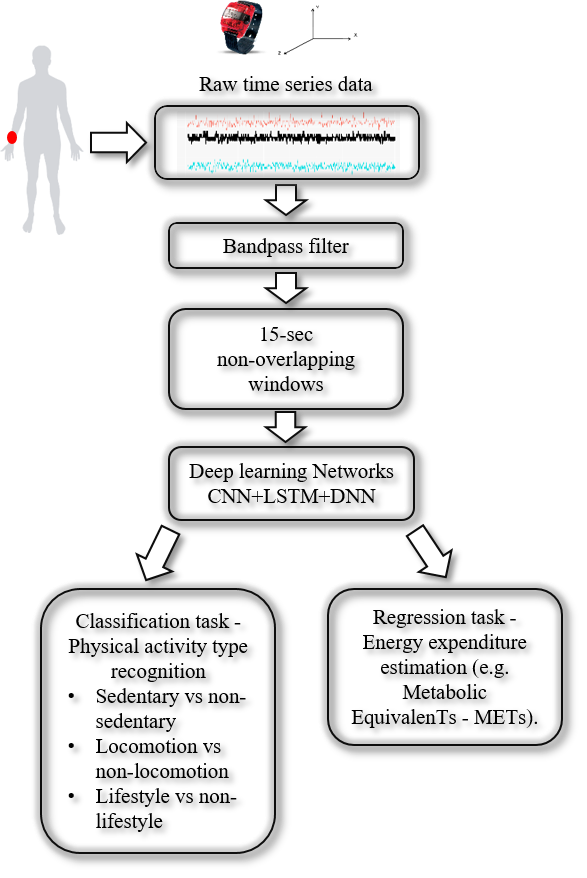}
	\caption{Accelerometry data processing overview. CNN refers to convolutional neural network; LSTM refers to Long-short-term memory (LSTM); and DNN refers to dense neural network.  }
	\label{fig:sys_overview}
\end{figure}

In this paper, we analyzed the raw accelerometry data that are collected from the wrist position in a laboratory setting. We filtered the data and split it into 15-seconds non-overlapping windows. These windows are used as an input to the deep learning network. The output of our models are the recognition of PA type and the estimation of EE. We hypothesize that the deep learning network is effective in extracting relevant features from the raw accelerometry data, predicting physical activity type, and estimating energy expenditure precisely from accelerometer data collected from the wrist. (Figure \ref{fig:sys_overview}) shows the overflow of the accelerometry data processing proposed in this work.

\section*{Methods}
\label{dtaa_coll}
\subsection*{Participants} One hundred and forty five participants (96 women and 49 men , aged 20-89 yrs) performed a battery of 33 typical daily activities in a standardized laboratory setting as listed in (Table \ref{tab:demographics}). Participants were community dwelling adults 20+ years old who can read and speak English language, were willing to undergo all testing procedures, and their weight was stable in the last three months (+/-5 lbs). (Table \ref{tab:demographics}) shows participants’ descriptive characteristics. Institutional Review Board at the University of Florida approved all study procedures, and all participants provided written informed consents before the study. 

\begin{table}[h]
	\centering
	\caption{List of the performed physical activities and their type}
	\label{tab:activities}
	\begin{tabular}{|l|c|c|c|}
		\hline
		&\multicolumn{3}{|c|}{\textbf{Activity type}}\\ \hline
		\textbf{Activity}&\textbf{Sedentary}&\textbf{Locomotion}&\textbf{Life-style}\\ \hline
		LEISURE WALK&No &Yes &No \\ \hline
		RAPID WALK&No &Yes &No \\ \hline
		LIGHT GARDENING&No &No  &Yes \\ \hline
		YARD WORK&No &No &Yes \\ \hline
		PREPARE SERVE MEAL&No &No &Yes \\ \hline
		DIGGING&No &No &Yes \\ \hline
		STRAIGHTENING UP DUSTING&No &No &Yes \\ \hline
		WASHING DISHES&No &No &Yes \\ \hline
		UNLOADING STORING DISHES&No &No &Yes \\ \hline
		WALKING AT RPE 1&No &Yes &No \\ \hline
		PERSONAL CARE&No &No &Yes \\ \hline
		DRESSING&No &No &Yes \\ \hline
		WALKING AT RPE 5&No &Yes &No \\ \hline
		SWEEPING&No &No &Yes \\ \hline
		VACUUMING&No &No &Yes \\ \hline
		STAIR DESCENT&No &Yes &No \\ \hline
		STAIR ASCENT&No &Yes &No \\ \hline
		TRASH REMOVAL&No &No &Yes \\ \hline
		REPLACING SHEETS ON A BED&No &No &Yes \\ \hline
		STRETCHING YOGA*&No &No &No \\ \hline
		MOPPING&No &No &Yes \\ \hline
		LIGHT HOME MAINTENANCE&No &No &Yes \\ \hline
		COMPUTER WORK& Yes &No &No \\ \hline
		HEAVY LIFTING&No &No &Yes \\ \hline
		SHOPPING&No &No &Yes \\ \hline
		IRONING&No &No &Yes \\ \hline
		LAUNDRY WASHING&No &No &Yes \\ \hline
		STRENGTH EXERCISE LEG CURL*&No &No &No \\ \hline
		STRENGTH EXERCISE CHEST PRESS*&No &No &No \\ \hline
		STRENGTH EXERCISE LEG EXTENSION*&No &No &No \\ \hline
		TV WATCHING&Yes &No &No \\ \hline
		STANDING STILL&Yes &No &No \\ \hline
		WASHING WINDOWS&No &No &Yes \\ \hline
		\multicolumn{4}{|l|}{A total of 29 activities were considered for PA type recognition and 33 for EE estimation}\\ 
		\multicolumn{4}{|l|}{* Only considered for energy expenditure estimation }\\\hline 
	\end{tabular}
\end{table}
A tri-axial accelerometer (ActiGraph GT3X+) was worn on the right wrist. Additionally, a portable metabolic unit (Cosmed K4b2) was worn to measure metabolic intensity that was expressed as a relative metabolic equivalent (MET). Tasks were chosen because they mimic daily chores activities, common among most Americans, and they are consistent with average time spent in the 2010 American Time Use Survey \cite{36}. Participants performed the scripted activities over four separate visits (2-3 hours each). Each one of these visits was designed to reduce the burden on participants and fatigue associated with performing physical activities over long periods of time. Activities were performed from lowest to highest metabolic demand with a 5-10 minute period between each activity. Full list of inclusion/exclusion criteria and data collection reproducibility can be found in the articles published by our group \cite{knaggs2011metabolic,26}.

\begin{table}[h]
	\centering
	\caption{Participants descriptive characteristics}
	\label{tab:demographics}
	\begin{tabular}{|l|c|}
		\hline
		&\textbf{Total (n=145)}\\ \hline
		\textbf{Age (yr)}&58.8 (17.1)\\ \hline
		\textbf{Female, n(\%)}&96 (66.2)\\ \hline
		\textbf{BMI ($kg*m^-2$)}&26.5 (4.8)\\ \hline
		\textbf{Race/ethnicity, n}&\\
		\quad Non-Hispanic &142\\
		\quad Hispanic&3\\ \hline
		\multicolumn{2}{|l|}{Data are means and SD unless otherwise noted.}\\
		\multicolumn{2}{|l|}{BMI: Body Mass Index}\\\hline
	\end{tabular}
\end{table}


\subsection*{Instrumentation} Participants wore an ActiGraph GT3X+ monitor on their right wrist. Findings suggest that wearing the accelerometer on the non‐dominant or dominant wrist has no impact on physical activity assessment \cite{dieu2017physical}. The ActiGraph GT3X+ monitor is a tri-axial lightweight accelerometer that records accelerations in units of gravity (1 g) in perpendicular, anterior-posterior, and medio-lateral axes. Accelerometers were initialized to collect data at 100 Hz sampling rate. Also, participants wore a portable indirect calorimetry device, Cosmed K4b2 \cite{cosmed}, while performing the activities listed in Table 1. Before data collection, the oxygen ($O_2$) and carbon dioxide ($CO_2$) sensors were calibrated using a gas mixture sample of 16.0\% $O_2$ and 5.0\% $CO_2$ and room air calibration. The turbine flow meter was calibrated using a 3.0-L syringe. A flexible facemask was positioned over the participant’s mouth and nose and attached to the flow meter. Oxygen consumption ($VO_2$ = $mL.min^{-1}.kg^{-1}$) was measured breath-by-breath. The breath-by-breath data were smoothed with a 30-sec running average window. The ventilation of $O_2$ ($VO_2$) data were displayed using a customize application in Labview to locate when steady state oxygen consumption was reached. Steady state was defined as a plateau in oxygen consumption, which typically occurs after 2 min of the start of the activity. $VO_2$ was averaged over a 2-4 minute steady state window. Data were expressed as METs after dividing the $VO_2$ values by the traditional standard of 3.5 $mL.min^{-1}.kg^{-1}$ \cite{jette1990metabolic}.

\subsection*{Problem Formulation}
\label{model_building}
Activities were split into three binary classification tasks: i) sedentary vs non-sedentary; ii) locomotion vs non-locomotion and iii) lifestyle vs non-lifestyle. MET values were examined as a continuous variable. We extracted consecutive non-overlapping 15-seconds windows from the raw accelerometry data and used it as an input ($x$) to the deep learning networks. Previous studies used various window lengths, ranging from 0.1 seconds to 128 seconds \cite{huynh2005analyzing,stikic2008adl,pirttikangas2006feature,mannini2013activity,krause2003unsupervised}. Our choice of 15-second window provided acceptable results. Participants performed the tasks listed in Table 1 at a self-selected pace for 10 min (treadmill walking was done for 7 min). Splitting these periods into 15 seconds intervals resulted in non-overlapping windows, and ensured that there is only one label for each window.	 The output of the model is a binary score $\hat{y}=(0, 1)$ in case of the classification tasks, and a continuous value in case of the regression task. We used binary cross-entropy loss function to measure the dissimilarity between the predicted ($\hat{y}$) and the ground truth ($y$) values, and optimized using Adam optimizer as shown in (Equation 1).

\begin{equation}
  \Lb=1/N\sum_{i=1}^{N}\lbrack y_i\log{\hat{y_i}} + (1-y_i)\log(1-\hat{y_i})\rbrack
\end{equation}

where $\Lb$ is the loss function, ($\hat{y}$) is the prediction, ($y$) is the ground truth; \textit{i} is the index of the raw accelerometry data window; and N is the number of samples.

\subsection*{Deep learning Model Architecture}
\begin{figure}[h]
	\centering
	\includegraphics[scale=0.65]{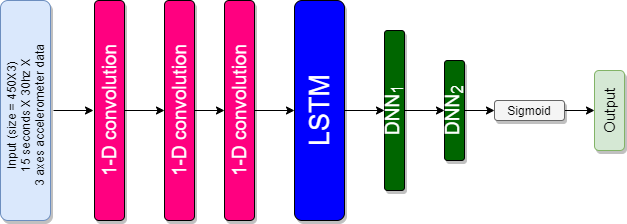}
	\caption{The architecture of the deep learning network.}
	\label{fig:archi}
\end{figure}

The deep learning network consists of multiple layers of neural networks where the processed output of one layer acts as the input to the next layer and so on. As any machine learning algorithm, the goal is to map the input to the target (labels). Technically, this is done through multiple data transformations (called layers) that are parameterized by numbers (called weights). Learning in this sense means finding the right weights to map the input to the target. Our network architecture comprised of multiple layers: input layer, convolutional neural network (CNN) layers, Long-short- term memory recurrent (LSTM) layer, and a classifier (dense neural network). The input layer is the raw accelerometry data split into 15-second non-overlapping windows. CNN is used to extract spatial features from the raw data. The LSTM layer is used to extract time features from the data. LSTM is a recurrent neural network (RNN) that includes a memory to carry information across timestamps, which can prevent the vanishing gradient problem \cite{hochreiter1998vanishing} - preventing older information from vanishing progressively throughout processing, which is important in time series analysis as in the case of accelerometry data \cite{franccois2017deep}. The output of LSTM is basically a feature map that is used by the classifier to recognize the PA type and estimate EE. It is worth mentioning that representations learned by CNN and LSTM are generic and reusable - useful regardless of the analyzed data. However, the classifier get rid of the spatial and temporal notions in the data resulting in representations that are specific to the classes in the problem. It will contain information about the probability of classes. We used binary cross-entropy loss function to measure how the classifier’s output is far from the expected one. Then the loss score is used as a feedback signal to update the weights of the network using Adam optimizer. In summary, our network comprised of 3 CNN layers, 1 LSTM layer, 2 layers of dense neural network, sigmoid activation, binary cross-entropy loss function, and Adam optimizer. (Figure \ref{fig:archi}) shows the architecture of the deep learning network.

\subsection*{Model Training}
Our dataset consisted of 145 participants included in the PA type recognition and 141 included in the EE estimation. Though the majority of accelerometry data were collected at 100 Hz frequency, a few cases were collected at 30 and 80 Hz frequencies. For the sake of consistency, we down sampled the frequencies to 30 hz using scipy python library. Then, we split the three-axis data into 15-seconds samples with 30 hz frequency, i.e., every sample had a shape of 450X3. As elucidated earlier, we categorized the PAs into three categories: sedentary, locomotion and lifestyle. We built three binary classification models and evaluated the performance of each one of them. All the 145 participants were randomly distributed into 10 batches, 9 batches of 15 participants each, and 1 batch of 10 participants. We used 10 fold nested cross validation to report the results on PA type classification with two loops. The outer loop considering every batch as test set and the inner loop with 9 iterations making every batch other than the test batch as validation set. In total there were 90 runs in our nested cross validation approach. 

There were more samples of lifestyle class compared to other classes making our dataset imbalanced. In the model training phase, we used weight balancing to assign higher weights to the minority class. This helped in preventing our binary models from becoming biased to the majority class. In the validation phase, we down-sampled the majority class, by randomly selecting samples, to make it equal to the minority class.

We used python 3.7 and keras deep learning library for our approach. The CNN-LSTM model is trained for 50 epochs. We used Adam optimizer and earlystopping callback with patience 5. The model is trained on Nvidia-Tesla K80 gpu for training. (Table \ref{tab:params}) provides technical information about the parameters and output sizes of each layer. For reproducibility purposes, we uploaded the code to our GitHub repository \cite{github} with a step-by-step manual explaining how to use them.

\begin{table}[h]
	\centering
	\caption{Parameters of the deep learning networks.}
	\label{tab:params}
	\begin{tabular}{|l|l|c|}
		\hline
		\textbf{Layer (type)}    & \textbf{Output Shape}  & \textbf{Param \#} \\ \hline
		$input_1$ (Input Layer)&[(None, 450, 3)]&0\\ \hline
		$conv1d$ (Conv1D)&[(None, 450, 16)]&400\\ \hline
		$conv1d_1$ (Conv1D)&[(None, 450, 32)]&4128\\ \hline
		$conv1d_2$ (Conv1D)&[(None, 450, 64)]&16448\\ \hline
		$lstm$ (LSTM)&[(None, 50)]&23000\\ \hline
		$dense$ (Dense)&[(None, 10)]&510\\ \hline
		$dense_1$ (Dense)&[(None, 1)]&11\\ \hline
		\multicolumn{3}{|l|}{Total params: 44,497}\\ 
		\multicolumn{3}{|l|}{Trainable params: 44,497}\\ 
		\multicolumn{3}{|l|}{Non-trainable params: 0}\\	\hline
	\end{tabular}
\end{table}

\section*{Results}
\label{results}

Balanced classification accuracy, sensitivity, specificity, precision, F1 score and area under the curve (AUC) metrics were used to evaluate the performance of the classification tasks. The selection of the balanced classification accuracy and F1 score was due to the data imbalance. Root mean square error (RMSE) was used to evaluate the regression task. The upper part of (Table \ref{tab:metrics}) shows the performance metrics of the classification tasks, and the lower part shows the RMSE value for all activities. Each one of the values is a mean of the 10 fold nested cross validation as explained earlier.

\begin{table}[h]
	\centering
	\caption{Performance metrics of recognizing physical activity type and estimating energy expenditure using deep learning networks. Each value is the mean and standard deviation of the 10 fold nested cross validation.}
	\label{tab:metrics}
	\begin{tabular}{|l|c|c|c|}
		\hline
		\multicolumn{4}{|c|}{\textbf{Activity Type}} \\ \hline
		Metric&Sedentary&Locomotion&Lifestyle\\ \hline
		Balanced Accuracy&0.89 (0.03)&	0.86 (0.05)&	0.88 (0.03)\\ \hline
		F1 score& 0.82 (0.05)&	0.81 (0.07)&	0.95 (0.01)\\ \hline
		AUC&0.98 (0.01)& 0.95 (0.02)& 0.96 (0.02)\\ \hline
		Sensitivity (Recall)&0.80 (0.07)&	0.72 (0.09)&	0.97 (0.01) \\ \hline
		Specificity& 0.98 (0.01)&	0.99 (0.01)&	0.79 (0.06) \\ \hline
		Precision& 0.85 (0.07)&	0.93 (0.06)&	0.94 (0.02) \\ \hline  
		\multicolumn{4}{|c|}{\textbf{Energy Expenditure}} \\ \hline
		RMSE&\multicolumn{3}{|c|}{1.1 (0.13)}\\ \hline
		\multicolumn{4}{|l|}{All values are mean and standard deviation.}\\ \hline
	\end{tabular}
\end{table}

\begin{figure}[h]
	\centering
	\includegraphics[scale=0.54]{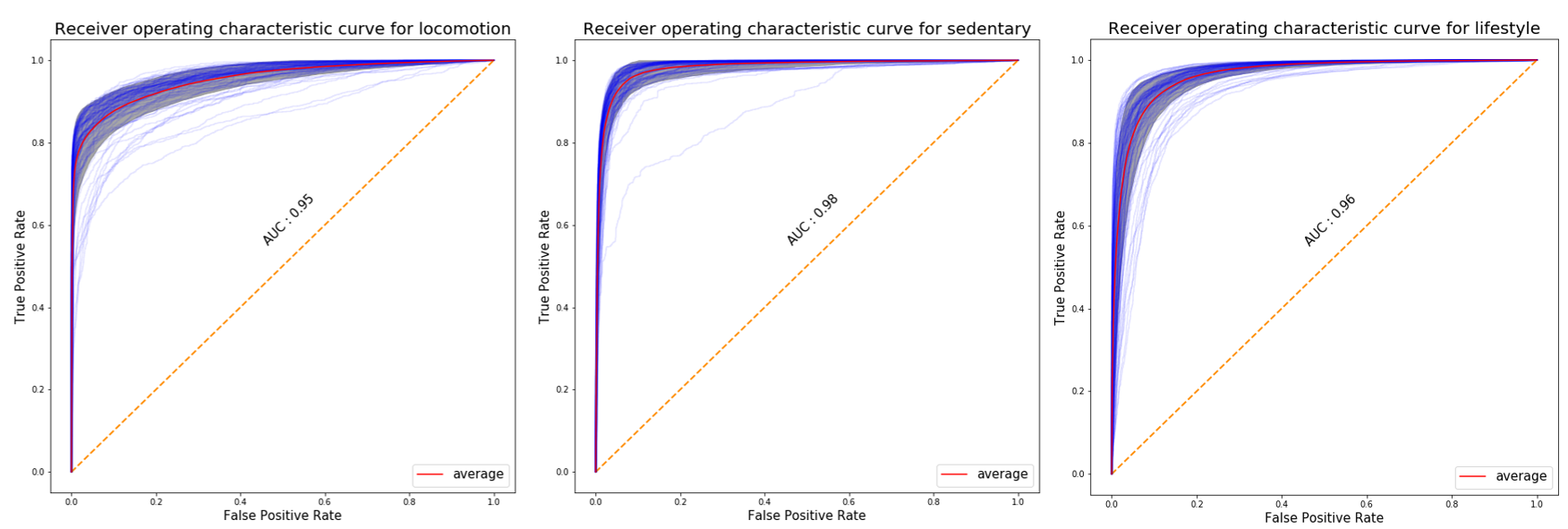}
	\caption{Receiver operating characteristic (ROC) curves for all the PA type classification tasks. Each one of the blue curves represents the curve from each run and the red curve is the mean over the 10 fold nested cross validation.}
	\label{fig:roc2}
\end{figure}

Figure \ref{fig:roc2} shows the receiver operator characteristic (ROC) curves for all the classification tasks. Each one of the blue curves represents the curve from each run and the red curve is the mean over the 10 fold nested cross validation. 

\section*{Discussion}
\label{diss}

The goal of the study was to show the effectiveness of the deep learning networks in extracting relevant features from the raw accelerometry data, predicting physical activity type, and estimating energy expenditure precisely from data collected from the wrist. We considered a deep learning network comprised of convolutional neural networks, long short-term networks and dense neural network for the classification and regression tasks. Results demonstrated that the deep learning models were relatively accurate at classifying physical activity types: i) sedentary vs non sedentary, ii) locomotion vs non-locomotion and iii) lifestyle vs non-lifestyle. Additionally, the models estimated the overall METs with reasonable accuracy — within +/- 1.1 METS or about 3.8 $ml$ of oxygen. 

The results of the deep learning models show relatively high performance in recognizing physical activity types. The lifestyle model achieved the highest F1 score (0.95), while sedentary and locomotion models were lower, 0.82 and 0.81, respectively. The high F1 score reflects the balance between precision and recall, which is indeed the case in lifestyle and sedentary recognition and a bit less in case of locomotion prediction. Lifestyle activities typically require more wrist involvement (i.e., ironing, trash removal) than other physical activity types. This can explain the high performance achieved by the lifestyle model. However, it seems from the results that less wrist involvement can lead to confusion in the deep learning model and reduce the performance as in the case of sedentary and locomotion activities. By looking descriptively at sensitivity scores, the lifestyle model seems to be confident in labeling lifestyle activities— neither under or over estimating the classifications. However, in case of sedentary, the model was less accurate and may over estimate activity level by misclassifying them as locomotion or lifestyle.

Comparing relevant literature results is an intricate endeavor because of the differences in the data collection environment and the variables that govern the study. There are numerous differences between studies that include: sample size, the demographic characteristics of participants, the number and diversity of the physical activities tested, type of accelerometer, body position, statistical and machine learning algorithms applied, the extracted statistical features, the window size, and the metrics measured to evaluate the overall performance. Given these differences, the results from work with a similar purpose are quite comparable. For example, Ellis et al. \cite{ellis2014random} built random forest models on data collected form the dominant wrist to predict physical activity type and estimate energy expenditure. The models were developed and tested on 40 (average age 35.8 years) participants. They obtained an average F1 score of 0.75 on 8 daily activities. Additionally, they obtained an RMSE value of 1.0 METs. Staudenmayer et al. \cite{staudenmayer2015methods} also used random forest to estimate energy expenditure and metabolic intensity of 19 physical activities from wrist accelerometry data. The models derived from a small young sample of 20 (24.1 years) estimated RMSE at 1.21 METs. When compared to others using machine learning approaches, the DL results from the current work are slightly better and reflect small RMSE MET differences between studies - Only +/- 0.20 METs.

Fewer studies have examined deep learning models to recognize physical activity type \cite{deep1,deep2,deep3,deep4}. The performance of the models \cite{deep3,deep4,deep2} built on the Opportunity dataset ranged between F1 score = [56.1 - 0.915] and an accuracy of 76.83\%. Additionally, the performance of the models built on Skoda dataset \cite{deep2,deep4} ranged between accuracy = [88.19 - 89.38]. Furthermore, the models \cite{deep1} built on WISDM and Dephnet were 98.23\% and 91.5\%, respectively. It should be noted that these studies have used publicly available data that contain activity labels, but not measures of metabolic intensity or energy expenditure (e.g. Opportunity\cite{opportunity} (multiple body positions, 3 participants), PAMAP2 (chest, arm and ankle positions, 9 participants)\cite{PAMAP2}, UCI daily and sports dataset (hip position, 30 participants)\cite{uci}, Skoda Mini Checkpoint (multiple body positions, 1 participant)\cite{skoda}, WISDM (hip position, 29 participants) \cite{WISDM}, and Daphnet Freezing of Gait Dataset (legs and hip positions, 10 participants)\cite{Daphnet}). They are also limited by a small number of participants, age-range being mostly $<$ 40 years, a low number and diversity of activity types, and most importantly lacking sufficient data from the wrist position. Given these substantial differences, the models presented here show relatively higher performance than others using DL approaches. Additionally, the current model may generalize better due to the high diversity of activities, wide age-span, gender and racial diversity and the larger number of participants enrolled.

A limitation of the current study is that data were collected in controlled lab settings, which is appropriate and a first step in evaluating positional differences \cite{keadle2019framework}. Collecting data in the free-living settings is more reflective of numerous transitions between activity types, but it is challenged by labeling the activity type. Another limitation is the consideration of window size, which was based on previous studies that extracted time- and frequency-domain features. This window size may not reflect the most appropriate size for deep learning networks. Additional simulation work should evaluate different window sizes for optimizing performance.

\section*{Conclusion}
\label{conc}

The goal of the study was to show the effectiveness of the deep learning networks in extracting relevant features from the raw accelerometry data, recognizing physical activity type, and estimating energy expenditure precisely from accelerometry data collected from the wrist. Deep learning networks comprising mainly of convolutional neural networks (CNN) and long-short-term memory (LSTM) demonstrated excellent performance in classifying broad activity types and estimating activity energy expenditure. As such, the spatial and temporal representations extracted from the deep learning models appear to be an effective substitution to the manual feature extraction. This knowledge is beneficial for the development of more accurate estimates of physical activity type and metabolic intensity for wrist accelerometers and mobile devices that use accelerometers (e.g. smart- watches) in both public and research arenas. 

\bibliographystyle{vancouver}
\bibliography{sample-base}

\end{document}